\documentclass{article}
\usepackage[margin=1in]{geometry}
\usepackage{parskip}
\usepackage{amsmath}
\usepackage{color}
\usepackage{natbib}
\usepackage[hidelinks]{hyperref}
\usepackage{amssymb}
\usepackage{booktabs}
\usepackage{longtable}
\usepackage{subcaption}
\usepackage{graphicx}
\graphicspath{{Figures/}}
\title{On relations of anisotropy and linear inhomogeneity using Backus average}
\author{
Md Abu Sayed\footnote{%
Department of Earth Sciences, Memorial University of Newfoundland,
{\tt m.abusayed.stu@gmail.com}}\,\,,
Theodore Stanoev\footnote{%
Department of Earth Sciences, Memorial University of Newfoundland,
{\tt theodore.stanoev@gmail.com}}
}
\date{}
\begin{document}
\maketitle
\begin{abstract}
The anisotropy of an equivalent medium resulting from the~\citet{Backus1962} average is induced by the vertical inhomogeneity among its constituent layers.
The velocity field of the constituent isotropic layers increases linearly with depth, which is assumed to be a good seismological description of sedimentary layers~\citep{Slotnick1959}.
We derive an analytical relationship between the anisotropy, characterized by the~\citet{Thomsen1986} parameters, and the linear inhomogeneity parameters, which forms a system of three equations for nine unknowns.
To obtain well-posedness, we constrain the problem by considering two seismological methods applied to field data.
We use the results from the two methods, for a particular region of interest, to assess the validity of the analytical relation.
\end{abstract}
\section{Introduction}

In this article, we parameterize an equivalent medium, resulting from the~\citet{Backus1962} average, with respect to linear inhomogeneity parameters of its constituent isotropic layers.
Sedimentary basins often show thin layers that induce seismic anisotropy.
The parametrization allows us to make a relationship between the anisotropy to layer inhomogeneity, which is, in seismological context, an important relation. 
We also assume the medium possesses linear inhomogeneity, which was first introduced by~\citet{Slotnick1959}.
The traveltime expression from a source to a receiver in a linearly inhomogeneous media is shown by~\citet{SlawinskiSlawinski1999}.

To obtain a solution from the analytical relationship between the anisotropy and the linear inhomogeneity parameters, 
we require to provide one of the inhomogeneity parameter.
We use two independent seismological methods, 1-D tomography and $ab$ model, to make the analytical relation to be well posed. 
The outcomes of the seismological methods are a pair of inhomogeneity parameters, 
using one of the parameter from each method allows us to obtain the solution of the analytical relation.
The resultant solution from the analytical relation then can be examined by comparing with the results of the seismological methods. 
\section{Equivalent medium parameterization}
We parameterize a transversely isotropic equivalent medium resulting from the~\citeauthor{Backus1962} average of thin, intrinsically homogeneous, isotropic layers, which we refer to as the~{\it\citeauthor{Backus1962} medium}.
We assume that a stack of such constituent layers is inhomogeneous and possesses a constant-velocity gradient that increases linearly with depth~\citep[see e.g.,][]{SlawinskiSlawinski1999}. 
Specifically, for both $P$ and $S$ waves, 
\begin{equation}
\label{eq:vPvS}
v_{P}\left(z\right) = a_{P} + b_{P}\,z
\quad{\rm and}\quad
v_{S}(z) = a_{S} + b_{S}\,z
\,,
\end{equation}
where~$a_{P,S}$ are the wavespeeds at the top of the medium,~$b_{P,S}$ are positive velocity-gradient constants, and $z$ is the depth.
\subsection{\citeauthor{Backus1962} average}
Following the definition of~\citet[Section~3]{Backus1962}, the average of a function $f(z)$ of width $\ell'$ is the moving average given by
\begin{equation}
\label{eq:BackusAverage}
\overline{f}(z)
=
\int\limits_{-\infty}^{\infty}
w\left(\zeta - z\right)
f\left(\zeta\right)
{\rm d}\zeta
\,,
\end{equation}
where the properties of the weighting function are
\begin{equation*}
w\left(z\right)
\geqslant
0
\,,\quad
w\left(\pm\infty\right) 
= 
0
\,,\quad
\int\limits_{-\infty}^{\infty}
w\left(z\right){\rm d}z
= 
1
\,,\quad
\int\limits_{-\infty}^{\infty}
z\,w\left(z\right){\rm d}z
=
0
\,,\quad
\int\limits_{-\infty}^{\infty}
z^{2}\,w\left(z\right){\rm d}z
=
\left(\ell'\right)^{2}
\,.
\end{equation*}
The result of performing average~\eqref{eq:BackusAverage} on isotropic layers results is a homogeneous TI medium, where the corresponding elasticity parameters, which are referred to as {\it Backus parameters}, are
\begin{subequations}
\begin{align}
\label{eq:cTI1111}
c^{\overline{\rm TI}}_{1111} 
&= 
\overline{
	\left(\dfrac{c_{1111}-2\,c_{2323}}{c_{1111}}\right)
}^{\,2}\,\,
\overline{\left(\dfrac{1}{c_{1111}}\right)}^{\,-1} +
\overline{
	\left(
		\dfrac{
			4\left(c_{1111} - c_{2323}\right)c_{2323}
		}{
			c_{1111}
		}
	\right)
}
\,,
\\
\label{eq:cTI1122}
c^{\overline{\rm TI}}_{1122} 
&= 
\overline{
	\left(\dfrac{c_{1111}-2\,c_{2323}}{c_{1111}}\right)
}^{\,2}\,\,
\overline{\left(\dfrac{1}{c_{1111}}\right)}^{\,-1} +
\overline{
	\left(
		\dfrac{
			2\left(c_{1111} - 2\,c_{2323}\right)c_{2323}
		}{
			c_{1111}
		}
	\right)
}
\,,
\\
\label{eq:cTI1133}
c^{\overline{\rm TI}}_{1133}
&= 
\overline{\left(\dfrac{c_{1111}-2\,c_{2323}}{c_{1111}}\right)}\,\,
\overline{\left(\dfrac{1}{c_{1111}}\right)}^{\,-1}
\,,
\\
\label{eq:cTI1212}
c^{\overline{\rm TI}}_{1212}
&=
\overline{c_{2323}}
\,,
\\
\label{eq:cTI2323}
c^{\overline{\rm TI}}_{2323} 
&=
\overline{\left(\dfrac{1}{c_{2323}}\right)}^{\,-1}
\,,
\\
\label{eq:cTI3333}
c^{\overline{\rm TI}}_{3333} 
&=
\overline{\left(\dfrac{1}{c_{1111}}\right)}^{\,-1}
\,.
\end{align}
\end{subequations}

Since the weighting function, $w$\,, in integral~\eqref{eq:BackusAverage}, is continuous and symmetric, the~\citeauthor{Backus1962} average may be written as a weighted average~\citep[e.g.,][Section~4.2.2]{Slawinski2018}.
Herein, the~\citeauthor{Backus1962} average is weighted by layer thickness.
For density-scaled VSP measurements, $v_{P} = \sqrt{c_{1111}}$ and $v_{S} = \sqrt{c_{2323}}$\,; this allows for a reparameterization of parameters~\eqref{eq:cTI1111}--\eqref{eq:cTI3333} in terms of linear-inhomogeneity parameters~\eqref{eq:vPvS}.
For example, Backus parameter~\eqref{eq:cTI1212} may be rewritten as
\begin{align}
\nonumber
c^{\overline{\rm TI}}_{1212}
=
\overline{c_{2323}}
=
\frac{1}{h_{2}-h_{1}}
\int\limits_{h_{1}}^{h_{2}}
c_{2323}\,
{\rm d}z
&=
\frac{1}{h_{2}-h_{1}}
\int\limits_{h_{1}}^{h_{2}}
\left(a_{S} + b_{S}\,z\right)^{2}
{\rm d}z
\\
\label{eq:AnIn_cTI1212}
&=
\frac{1}{3}
\left(
	3\,a_{S}^{2} + 
	3\,a_{S}\,b_{S}\left(h_{1} + h_{2}\right) + 
	b_{S}^{2}\left(h_{1}^{2} + h_{1}\,h_{2} + h_{2}^{2}\right)
\right)
\,.
\end{align}

For Backus parameter~\eqref{eq:cTI1111}, we begin with the first term, where
\begin{equation}
\label{eq:AnIn_cTI1111_t1}
\overline{\left(\dfrac{c_{1111}-2\,c_{2323}}{c_{1111}}\right)}
=
\dfrac{1}{h_{2}-h_{1}}
\int\limits_{h_{1}}^{h_{2}}
\left(
	1
	-
	2\,\dfrac{
		{v_{S}}^{2}
	}{
		{v_{P}}^{2}
	}
\right)
{\rm d}z
=
\dfrac{1}{h_{2}-h_{1}}
\int\limits_{h_{1}}^{h_{2}}
\left(
	1
	-
	2\,\dfrac{
		\left(a_{S} + b_{S}\,z\right)^{2}
	}{
		\left(a_{P} + b_{P}\,z\right)^{2}
	}
\right)
{\rm d}z
=
1-\dfrac{2\,I_{1}}{h_{2} - h_{1}}
\end{equation}
and
\begin{align}
\nonumber
I_{1}
=
\int\limits_{h_{1}}^{h_{2}}
\dfrac{
	\left(a_{S}+b_{S}\,z\right)^{2}
}{
	\left(a_{P}+b_{P}\,z\right)^{2}
}\,
{\rm d}z
&=
\dfrac{h_{2}\,{b_{S}}^2}{{b_{P}}^2} -
\dfrac{h_{1}\,{b_{S}}^2}{{b_{P}}^2}
\\
\nonumber
&+
\dfrac{
	\ln\left(a_{P} + h_{1}\,b_{P}\right)
	\left(
		2\,a_{P}\,{b_{S}}^{2} -
		2\,a_{S}\,b_{P}\,b_{S}
	\right)
}{
	{b_{P}}^{3}
} 
-
\dfrac{
	\ln\left(a_{P}+h_{2}\,b_{P}\right)
	\left(
		2\,a_{P}\,{b_{S}}^{2} -
		2\,a_{S}\,b_{P}\,b_{S}
	\right)
}{
	{b_{P}}^3
}
\\
\label{eq:I1}
&+
\dfrac{
	{a_{P}}^{2}\,{b_{S}}^2 -
	2\,a_{P}\,a_{S}\,b_{P}\,b_{S} +
	{a_{S}}^2\,{b_{P}}^2
}{
	b_{P}
	\left(
		h_{1}\,{b_{P}}^{3} +
		a_{p}\,{b_{P}}^{2}
	\right)
}
-
\dfrac{
	{a_{P}}^2\,{b_{S}}^{2} -
	2\,a_{P}\,a_{S}\,b_{P}\,b_{S} +
	{a_{S}}^2\,{b_{P}}^2
}{
	b_{P}
	\left(h_{2}\,{b_{P}}^{3} + a_{P}\,{b_{P}}^{2}\right)
}
\,.
\end{align}
For the second term in parameter~\eqref{eq:cTI1111}, 
\begin{equation*}
\overline{\left(\dfrac{1}{c_{1111}}\right)}^{\,-1}
=
\overline{\left(\dfrac{1}{v_{P}^{2}}\right)}^{\,-1}
=
\left(
	\dfrac{1}{h_{2}-h_{1}}
	\int\limits_{h_{1}}^{h_{2}}
	\dfrac{1}{\left(a_{P} + b_{P}\,z\right)^{2}}\,
	\mathrm{d}z
\right)^{\!\!\!-1}
=
\left(h_{2} - h_{1}\right)
\left(\,\,
	\int\limits_{h_{1}}^{h_{2}}
	\left(a_{P} + b_{P}\,z\right)^{-2}\,
	\mathrm{d}z
\right)^{\!\!\!-1}
\,.
\end{equation*}
Substituting $u$ for~$a_{P}+b_{P}z$\,, which gives $\mathrm{d}z = \dfrac{\mathrm{d}u}{b_{P}}$\,, and changing the limits, $h_{1}\to a_{P} + b_{P}\,h_{1}$ and $h_{2}\to a_{P} + b_{P}\,h_{2}$\,, we obtain
\begin{align}
\nonumber
\overline{\left(\dfrac{1}{c_{1111}}\right)}^{\,-1}
&=
\left(h_{2} - h_{1}\right)
\left(\,\,
	\int\limits_{a_{P} + b_{P}\,h_{1}}^{a_{P} + b_{P}\,h_{2}} 
	u^{-2} 
	\dfrac{\mathrm{d}u}{b_{P}}
\right)^{\!\!\!-1}
=
\left(h_{2} - h_{1}\right)
\left(
	\left.
		\dfrac{-u^{-1}}{b_{P}}
	\right|_{a_{P} + b_{P}\,h_{1}}^{a_{P} + b_{P}\,h_{2}} 
\right)^{\!\!\!-1}
\\
\nonumber
&=
\left(h_{2} - h_{1}\right)
\left(-b_{P}\right)
\left(
	\left(a_{P} + b_{P}\,h_{2}\right)^{-1} -
	\left(a_{P} + b_{P}\,h_{1}\right)^{-1}
\right)^{-1}
\\
\label{eq:AnIn_cTI3333}
&=
\left(a_{P} + b_{P}\,h_{1}\right)
\left(a_{P} + b_{P}\,h_{2}\right)
\,.
\end{align}

For the last term in parameter~\eqref{eq:cTI1111},
\begin{equation}
\label{eq:AnIn_cTI1111_t3}
\overline{
	\left(
		\dfrac{
			4\left(c_{1111} - c_{2323}\right)c_{2323}
		}{
			c_{1111}
		}
	\right)
}
=
\dfrac{4}{h_{2}-h_{1}}
\left(\,\,
	\int\limits_{h_{1}}^{h_{2}}
	\left(a_{S} + b_{S}\,z\right)^{2}
	{\rm d}z
	-
	\int\limits_{h_{1}}^{h_{2}}
	\dfrac{
		\left(a_{S} + b_{S}\,z\right)^{4}
	}{
		\left(a_{P} + b_{P}\,z\right)^{2}
	}\,
	{\rm d}z
\right)
=
\dfrac{4}{h_{2}-h_{1}}\left(I_{2} - I_{3}\right)
\,,
\end{equation}
where
\begin{equation}
\label{eq:I2}
I_{2}
=
\int\limits_{h_{1}}^{h_{2}}
\left(a_{S} + b_{S}\,z\right)^{2}\,
{\rm d}z
=
\frac{1}{3}
\left(h_{2} - h_{1}\right)
\left(
	3\,a_{S}{}^2 +
	3\,a_{S}\,b_{S}\left(h_{1} + h_{2}\right) + 
	b_{S}{}^{2}
	\left(h_{1}^{2} + h_{1}\,h_{2} + h_{2}^{2}\right)
\right)
\end{equation}
and
\begin{align}
\nonumber
I_{3}
=
\int\limits_{h_{1}}^{h_{2}}
\dfrac{
	\left(a_{S} + b_{S}\,z\right)^{4}
}{
	\left(a_{P} + b_{P}\,z\right)^{2}
}\,
{\rm d}z
&=
h_{2}
\left(
	\dfrac{
		2\,a_{p}
		\left(
			\dfrac{2\,a_{P}\,{b_{S}}^{4}}{{b_{P}}^{3}} -
			\dfrac{4\,a_{S}\,{b_{S}}^{3}}{{b_{P}}^{2}}
		\right)
	}{
		b_{P}
	} 
	-
	\dfrac{{a_{P}}^{2}\,{b_{S}}^{4}}{{b_{P}}^{4}} +
	\dfrac{6\,{a_{S}}^{2}\,{b_{S}}^{2}}{{b_{P}}^{2}}
\right)
\\\nonumber
&-
h_{1}
\left(
	\dfrac{
		2\,a_{P}
		\left(
			\dfrac{2\,a_{P}\,{b_{S}}^{4}}{{b_{P}}^{3}} -
			\dfrac{4\,a_{S}\,{b_{S}}^{3}}{{b_{P}}^{2}}
		\right)
	}{
		b_{P}
	} 
	-
	\dfrac{{a_{P}}^{2}\,{b_{S}}^{4}}{{b_{P}}^{4}} +
	\dfrac{6\,{a_{S}}^{2}\,{b_{S}}^{2}}{{b_{P}}^{2}}
\right)
\\\nonumber
&+
{h_{1}}^2
\left(
	\dfrac{a_{P}\,{b_{S}}^{4}}{{b_{P}}^{3}} -
	\dfrac{2\,a_{S}\,{b_{S}}^{3}}{{b_{P}}^{2}}
\right)
-
{h_{2}}^2
\left(
	\dfrac{a_{P}\,{b_{S}}^{4}}{{b_{P}}^{3}} -
	\dfrac{2\,a_{S}\,{b_{S}}^{3}}{{b_{P}}^{2}}
\right)
\\\nonumber
&+
\dfrac{
	\ln\left(a_{P} + h_{1}\,b_{P}\right)
	\left(
		4\,{a_{P}}^{3}\,{b_{S}}^{4} -
		12\,{a_{P}}^{2}\,a_{S}\,b_{P}\,{b_{S}}^{3} +
		12\,a_{P}\,{a_{S}}^{2}\,{b_{P}}^{2}\,{b_{S}}^{2} -
		4\,{a_{S}}^{3}\,{b_{P}}^{3}\,b_{S}
	\right)
}{
	{b_{P}}^5
}
\\\nonumber
&-
\dfrac{
	\ln\left(a_{P} + h_{2}\,b_{P}\right)
	\left(
		4\,{a_{P}}^{3}\,{b_{S}}^{4} -
		12\,{a_{P}}^{2}\,a_{S}\,b_{P}\,{b_{S}}^{3} +
		12\,a_{P}\,{a_{S}}^{2}\,{b_{P}}^{2}\,{b_{S}}^{2} -
		4\,{a_{S}}^{3}\,{b_{P}}^{3}\,b_{S}
	\right)
}{
	{b_{P}}^5
}
\\\nonumber
&-
\dfrac{{h_{1}}^{3}\,{b_{S}}^{4}}{3\,{b_{P}}^{2}} +
\dfrac{{h_{2}}^{3}\,{b_{S}}^{4}}{3\,{b_{P}}^{2}}
\\\nonumber
&+
\dfrac{
	{a_{P}}^{4}\,{b_{S}}^{4} -
	4\,{a_{P}}^{3}\,a_{S}\,b_{P}\,{b_{S}}^{3} +
	6\,{a_{P}}^{2}\,{a_{S}}^{2}\,{b_{P}}^{2}\,{b_{S}}^{2} -
	4\,a_{P}\,{a_{S}}^{3}\,{b_{P}}^{3}\,b_{S} +
	{a_{S}}^{4}\,{b_{P}}^{4}
}{
	b_{P}
	\left(h_{1}\,{b_{P}}^{5} + a_{P}\,{b_{P}}^{4}\right)
}
\\\label{eq:I3}
&-
\dfrac{
	{a_{P}}^{4}\,{b_{S}}^{4} -
	4\,{a_{P}}^{3}\,a_{S}\,b_{P}\,{b_{S}}^{3} +
	6\,{a_{P}}^{2}\,{a_{S}}^{2}\,{b_{P}}^{2}\,{b_{S}}^{2} -
	4\,a_{P}\,{a_{S}}^{3}\,{b_{P}}^{3}\,b_{S} +
	{a_{S}}^{4}\,{b_{P}}^{4}
}{
	b_{P}
	\left(h_{2}\,{b_{P}}^{5} + a_{P}\,{b_{P}}^{4}\right)
}
\,.
\end{align}

The first two terms in parameter~\eqref{eq:cTI1122} are given by formul\ae~\eqref{eq:AnIn_cTI1111_t1} and~\eqref{eq:AnIn_cTI3333}, whereas the third term is
\begin{equation}
\label{eq:AnIn_cTI1122_t3}
\overline{
	\left(
		\dfrac{
			2\left(c_{1111} - 2\,c_{2323}\right)c_{2323}
		}{
			c_{1111}
		}
	\right)
}
=
\frac{2}{h_{2}-h_{1}}
\left(\,\,
	\int\limits_{h_{1}}^{h_{2}}
	\left(a_{S}+b_{S}\,z\right)^{2}
	{\rm d}z
	-
	2
	\int\limits_{h_{1}}^{h_{2}}
	\frac{
		\left(a_{S} + b_{S}\,z\right)^{4}
	}{
		\left(a_{P} + b_{P}\,z\right)^{2}
	}\,
	{\rm d}z
\right)
=
\frac{2}{h_{2}-h_{1}}
\left(I_{2} - 2\,I_{3}\right)
\,,
\end{equation}
where $I_{2}$ and $I_{3}$ are given by integration constants~\eqref{eq:I2} and~\eqref{eq:I3}.
Finally, in a manner similar to obtaining the second term in parameter~\eqref{eq:cTI1111}, we use $u$ substitution and change limits of integration to obtain the term in parameter~\eqref{eq:cTI2323}, where
\begin{align}
\nonumber
\overline{\left(\dfrac{1}{c_{2323}}\right)}^{\,-1}
&=
\left(
	\frac{1}{h_{2}-h_{1}}
	\int\limits_{h_{1}}^{h_{2}}
	\dfrac{1}{\left(a_{S} + b_{S}\,z\right)^{2}}\,
	{\rm d}z
\right)^{\!\!\!-1}
=
\left(h_{2} - h_{1}\right)
\left(\,\,
	\int\limits_{h_{1}}^{h_{2}}
	\left(a_{S} + b_{S}\,z\right)^{-2}\,
	\mathrm{d}z
\right)^{\!\!\!-1}
\\
\label{eq:AnIn_cTI2323}
&=
\left(a_{S} + b_{S}\,h_{1}\right)
\left(a_{S} + b_{S}\,h_{2}\right)
\,.
\end{align}

Thus, using formul\ae~\eqref{eq:AnIn_cTI1212},~\eqref{eq:AnIn_cTI1111_t1},~\eqref{eq:AnIn_cTI3333},~\eqref{eq:AnIn_cTI1111_t3},~\eqref{eq:AnIn_cTI1122_t3},~\eqref{eq:AnIn_cTI2323}, along with integration constants~\eqref{eq:I1},~\eqref{eq:I2},~\eqref{eq:I3}, we may restate Backus parameters~\eqref{eq:cTI1111}--\eqref{eq:cTI3333} as
\begin{subequations}
\label{eq:cTI_abh1h2}
\begin{align}
\label{eq:cTI1111_abh1h2}
c^{\overline{\rm TI}}_{1111}
\left(h_{1}\,,h_{2}\,,a_{S}\,,b_{S}\,,a_{P}\,,b_{P}\right)
&= 
\left(1-\dfrac{2\,I_{1}}{h_{2} - h_{1}}\right)^{2}
\left(a_{P} + b_{P}\,h_{1}\right)
\left(a_{P} + b_{P}\,h_{2}\right)
+
\frac{4}{h_{2}-h_{1}}
\left(I_{2} - I_{3}\right)
\,,
\\
\label{eq:cTI1122_abh1h2}
c^{\overline{\rm TI}}_{1122}
\left(h_{1}\,,h_{2}\,,a_{S}\,,b_{S}\,,a_{P}\,,b_{P}\right)
&= 
\left(1-\dfrac{2\,I_{1}}{h_{2} - h_{1}}\right)^{2}
\left(a_{P} + b_{P}\,h_{1}\right)
\left(a_{P} + b_{P}\,h_{2}\right)
+
\frac{2}{h_{2}-h_{1}}
\left(I_{2} - 2\,I_{3}\right)
\,,
\\
\label{eq:cTI1133_abh1h2}
c^{\overline{\rm TI}}_{1133}
\left(h_{1}\,,h_{2}\,,a_{S}\,,b_{S}\,,a_{P}\,,b_{P}\right)
&= 
\left(1-\dfrac{2\,I_{1}}{h_{2} - h_{1}}\right)
\left(a_{P} + b_{P}\,h_{1}\right)
\left(a_{P} + b_{P}\,h_{2}\right)
\,,
\\
\label{eq:cTI1212_abh1h2}
c^{\overline{\rm TI}}_{1212}
\left(h_{1}\,,h_{2}\,,a_{S}\,,b_{S}\right)
&=
\frac{1}{3}
\left(
	3\,a_{S}^{2} + 
	3\,a_{S}\,b_{S}\left(h_{1} + h_{2}\right) + 
	b_{S}^{2}\left(h_{1}^{2} + h_{1}\,h_{2} + h_{2}^{2}\right)
\right)
\,,
\\
\label{eq:cTI2323_abh1h2}
c^{\overline{\rm TI}}_{2323}
\left(h_{1}\,,h_{2}\,,a_{S}\,,b_{S}\right)
&=
\left(a_{S} + b_{S}\,h_{1}\right)
\left(a_{S} + b_{S}\,h_{2}\right)
\,,
\\
\label{eq:cTI3333_abh1h2}
c^{\overline{\rm TI}}_{3333}
\left(h_{1}\,,h_{2}\,,a_{P}\,,b_{P}\right)
&=
\left(a_{P} + b_{P}\,h_{1}\right)
\left(a_{P} + b_{P}\,h_{2}\right)
\,.
\end{align}
\end{subequations}
\subsection{Anisotropy parameters}
The anisotropy of any transversely isotropic medium may be described by the~\citet{Thomsen1986} parameters.
Thus, using Backus parameters~\eqref{eq:cTI_abh1h2}, we define
\begin{subequations}
\label{eq:ThomsenParam_linInh}
\begin{align}
\label{eq:gamma_linInh}
\gamma
=
\gamma\left(h_{1}\,,h_{2}\,,a_{S}\,,b_{S}\right)
&:=
\frac{
	c^{\overline{\rm TI}}_{1212} - 
	c^{\overline{\rm TI}}_{2323}
}{
	2\,c^{\overline{\rm TI}}_{2323}
}
\,,
\\
\label{eq:delta_linInh}
\delta
=
\delta\left(h_{1}\,,h_{2}\,,a_{S}\,,b_{S}\,,a_{P}\,,b_{P}\right)
&:=
\frac{
	\left(
		c^{\overline{\rm TI}}_{1133} +
		c^{\overline{\rm TI}}_{2323}
	\right)^{2}
	-
	\left(
		c^{\overline{\rm TI}}_{3333} -
		c^{\overline{\rm TI}}_{2323}
	\right)^{2}
}{
	2\,c^{\overline{\rm TI}}_{3333}
	\left(
		c^{\overline{\rm TI}}_{3333} -
		c^{\overline{\rm TI}}_{2323}
	\right)
}
\,,
\\
\label{eq:epsilon_linInh}
\varepsilon
=
\varepsilon\left(h_{1}\,,h_{2}\,,a_{S}\,,b_{S}\,,a_{P}\,,b_{P}\right)
&:=
\frac{
	c^{\overline{\rm TI}}_{1111} - 
	c^{\overline{\rm TI}}_{3333}
}{
	2\,c^{\overline{\rm TI}}_{3333}
}
\,.
\end{align}
\end{subequations}

Explicitly, parameter~\eqref{eq:gamma_linInh} is
\begin{subequations}
\begin{equation}
\label{eq:gamma_abh1h2}
\gamma
= 
\gamma\left(h_{1}\,,h_{2}\,,a_{S}\,,b_{S}\right)
=
\frac{
	{b_{S}}^{2}\left(h_{1} - h_{2}\right)^{2}
}{
	6
	\left(a_{S} + b_{S}\,h_{1}\right)
	\left(a_{S} + b_{S}\,h_{2}\right)
}
\,.
\end{equation}
For parameter~\eqref{eq:delta_linInh},
\begin{align}
\nonumber
\delta
&= 
\delta
\left(h_{1}\,,h_{2}\,,a_{S}\,,b_{S}\,,a_{P}\,,b_{P}\right)
\\
\label{eq:delta_abh1h2}
&=
\frac{
	2\,{\delta}_{k1}
	\left(
		-b_{P}
		\left(h_{1} - h_{2}\right)
		\left(
			2\,a_{P} + b_{P}\left(h_{1} + h_{2}\right)
		\right)
		+
		{\delta}_{k2}
	\right)
	\left(
		b_{P}
		\left(h_{1} - h_{2}\right)
		{\delta}_{k3}
		+
		{\delta}_{k1}\,
		{\delta}_{k2}
	\right)
}{
	{b_{P}}^6
	\left(a_{P} + b_{P}\,h_{1}\right)
	\left(h_{1} - h_{2}\right)^{2}
	\left(a_{P} + b_{P}\,h_{2}\right)
	\left(
		{a_{P}}^2 +
		{\delta}_{k4} +
		a_{P}\,b_{P}\left(h_{1} + h_{2}\right) -
		a_{S}\,b_{S}\left(h_{1} + h_{2}\right)
	\right)
}
\,,
\end{align}
where
\begin{align*}
{\delta}_{k1}
&=
b_{S}
\left(-a_{S}\,b_{P} + a_{P}\,b_{S}\right)
\,,
\\
{\delta}_{k2}
&= 
2\left(a_{P} + b_{P}\,h_{1}\right)
\left(a_{P} + b_{P}\,h_{2}\right)
\ln\left(
	\frac{a_{P} + b_{P}\,h_{1}}{a_{P} + b_{P}\,h_{2}}
\right)
\,,
\\
{\delta}_{k3}
&=
\left(
	{a_{P}}^{2}\left({b_{P}}^2 - 2\,{b_{S}}^2\right) +
	{b_{P}}^{2}\left(
		{\delta}_{k4}
	\right)
	+
	a_{P}\,b_{P}\left(
		2\,a_{S}\,b_{S} +
		\left(b_{P} - b_{S}\right)
		\left(b_{P} + b_{S}\right)
		\left(h_{1} + h_{2}\right)
	\right)
\right)
\,,
\\
{\delta}_{k4}
&=
-a_{S}^2 + 
\left(b_{P} - b_{S}\right)
\left(b_{P} + b_{S}\right)\,h_{1}\,h_{2}
\,.
\end{align*}
For parameter~\eqref{eq:epsilon_linInh},
\begin{align}
\label{eq:epsilon_abh1h2}
\varepsilon
&=
\varepsilon
\left(h_{1}\,,h_{2}\,,a_{S}\,,b_{S}\,,a_{P}\,,b_{P}\right)
=
\frac{
	2\,b_{S}
	\left(
		{b_{P}}^{3}
		\left(h_{1} - h_{2}\right)^{2}
		\left(
			-6\,{a_{P}}^{2}\,b_{P}\,b_{S} +
			b_{P}
			\left(
				{\varepsilon}_{k1}
			\right) 
			+
			3\,a_{P}
			\left(
				{\varepsilon}_{k2}
			\right)
		\right)
		+
		\left({\varepsilon}_{k3}\right)
		\left({\varepsilon}_{k4}\right)
	\right)
}{
	3\,{b_{P}}^{6}\left(a_{P} + b_{P}\,h_{1}\right)
	\left(h_{1} - h_{2}\right)^{2}
	\left(a_{P} + b_{P}\,h_{2}\right)
}
\end{align}
\end{subequations}
where
\begin{align*}
{\varepsilon}_{k1}
&=
-12\,{a_{S}}^{2}\,b_{S} 
+
\left(b_{P} - b_{S}\right)
b_{S}
\left(b_{P} + b_{S}\right)
\left(h_{1} - h_{2}\right)^{2} 
+
3\,a_{S}
\left({b_{P}}^{2} - 2\,{b_{S}}^{2}\right)
\left(h_{1} + h_{2}\right)
\,,
\\
{\varepsilon}_{k2}
&=
2\,a_{S}
\left({b_{P}}^{2} + 2\,{b_{S}}^{2}\right)
+
b_{S}
\left(-{b_{P}}^{2} + 2\,{b_{S}}^{2}\right)
\left(h_{1} + h_{2}\right)
\,,
\\
{\varepsilon}_{k3}
&=
6
\left(a_{S}\,b_{P} - a_{P}\,b_{S}\right)
\left(a_{P} + b_{P}\,h_{1}\right)
\left(a_{P} + b_{P}\,h_{2}\right)
\ln\left(
	\frac{a_{P} + b_{P}\,h_{1}}{a_{P} + b_{P}\,h_{2}}
\right)
\,,
\\
{\varepsilon}_{k4}
&=
-b_{P}
\left({b_{P}}^{2} - 2\,{b_{S}}^{2}\right)
\left(h_{1} - h_{2}\right)
+
2\,b_{S}\left(-a_{S}\,b_{P} + a_{P}\,b_{S}\right)
\ln\left(
	\frac{a_{P} + b_{P}\,h_{2}}{a_{P} + b_{P}\,h_{1}}
\right)
\,.
\end{align*}
\subsection{Total differentials of anisotropy parameters}
To quantify the uncertainty of expressions~\eqref{eq:gamma_abh1h2}--\eqref{eq:epsilon_abh1h2}, i.e., the sensitivity to changes in model parameters, we require the total differential.
We demonstrate this, we differentiate $\gamma$ with respect to each of its coordinates directions to obtain its linear functional
\begin{subequations}
\begin{equation}
\label{eq:gammaDifferential}
{\rm d}\gamma
=
\left(
	\frac{\partial\,\gamma}{\partial h_{1}}
\right)
{\rm d}h_{1} 
+
\left(
	\frac{\partial\,\gamma}{\partial h_{2}}
\right)
{\rm d}h_{2} 
+
\left(
	\frac{\partial\,\gamma}{\partial a_{S}}
\right)
{\rm d}a_{S} 
+
\left(
	\frac{\partial\,\gamma}{\partial b_{S}}
\right)
{\rm d}b_{S}
\,,
\end{equation}
where
\begin{align*}
\frac{\partial\,\gamma}{\partial h_{1}}
&=
\frac{
	{b_{S}}^{2}
	\left(h_{1} - h_{2}\right)
	\left(
		2\,a_{S} + 
		b_{S}\left(h_{1} + h_{2}\right)
	\right)
}{
	6
	\left(a_{S} + b_{S}\,h_{1}\right)^{2}
	\left(a_{S} + b_{S}\,h_{2}\right)
}
\,,
\\
\frac{\partial\,\gamma}{\partial h_{2}}
&=
-\frac{
	{b_{S}}^{2}
	\left(h_{1} - h_{2}\right)
	\left(
		2\,a_{S} + 
		b_{S}\left(h_{1} + h_{2}\right)
	\right)
}{
	6
	\left(a_{S} + b_{S}\,h_{1}\right)
	\left(a_{S} + b_{S}\,h_{2}\right)^{2}
}
\,,
\\
\frac{\partial\,\gamma}{\partial a_{S}}
&=
-\frac{
	{b_{S}}^{2}
	\left(h_{1} - h_{2}\right)^{2}
	\left(
		2\,a_{S} + 
		b_{S}\left(h_{1} + h_{2}\right)
	\right)
}{
	6
	\left(a_{S} + b_{S}\,h_{1}\right)^{2}
	\left(a_{S} + b_{S}\,h_{2}\right)^{2}
}
\,,
\\
\frac{\partial\,\gamma}{\partial b_{S}}
&=
\frac{
	a_{S}\,b_{S}
	\left(h_{1} - h_{2}\right)^{2}
	\left(
		2\,a_{S} + 
		b_{S}\left(h_{1} + h_{2}\right)
	\right)
}{
	6
	\left(a_{S} + b_{S}\,h_{1}\right)^{2}
	\left(a_{S} + b_{S}\,h_{2}\right)^{2}
}
\,.
\end{align*}
We may perform similar operations on $\delta$ and $\varepsilon$ to obtain
\begin{equation}
\label{eq:deltaDifferential}
{\rm d}\delta
=
\left(\frac{\partial\,\delta}{\partial h_{1}}\right){\rm d}h_{1} +
\left(\frac{\partial\,\delta}{\partial h_{2}}\right){\rm d}h_{2} +
\left(\frac{\partial\,\delta}{\partial a_{S}}\right){\rm d}a_{S} +
\left(\frac{\partial\,\delta}{\partial b_{S}}\right){\rm d}b_{S} +
\left(\frac{\partial\,\delta}{\partial a_{P}}\right){\rm d}a_{P} +
\left(\frac{\partial\,\delta}{\partial b_{P}}\right){\rm d}b_{P}
\end{equation}
and
\begin{equation}
\label{eq:epsilonDifferential}
{\rm d}\varepsilon
=
\left(\frac{\partial\,\varepsilon}{\partial h_{1}}\right){\rm d}h_{1} +
\left(\frac{\partial\,\varepsilon}{\partial h_{2}}\right){\rm d}h_{2} +
\left(\frac{\partial\,\varepsilon}{\partial a_{S}}\right){\rm d}a_{S} +
\left(\frac{\partial\,\varepsilon}{\partial b_{S}}\right){\rm d}b_{S} +
\left(\frac{\partial\,\varepsilon}{\partial a_{P}}\right){\rm d}a_{P} +
\left(\frac{\partial\,\varepsilon}{\partial b_{P}}\right){\rm d}b_{P}
\,.
\end{equation}
\end{subequations}
We do not list the the resultant expressions since they would require half-a-dozen pages.
However, the expressions for the partial derivatives of differentials~\eqref{eq:deltaDifferential} and~\eqref{eq:epsilonDifferential} may be derived using a symbolic software, such as Mathematica.
\section{Methods for well-posedness}
\label{sec:Methods}
Expressions~\eqref{eq:gamma_abh1h2}--\eqref{eq:epsilon_abh1h2} form an ill-posed system of three equations with nine unknowns, where
\begin{equation}
\label{eq:AnalyticSystem}
\begin{cases}
\gamma
=
\gamma\left(h_{1},h_{2},a_{S},b_{S}\right)
\\
\delta
=
\delta\left(h_{1},h_{2},a_{S},b_{S},a_{P},b_{P}\right)
\\
\varepsilon
=
\varepsilon\left(h_{1},h_{2},a_{S},b_{S},a_{P},b_{P}\right)
\end{cases}
\,.
\end{equation}
We use differentials~\eqref{eq:gammaDifferential}--\eqref{eq:epsilonDifferential} to form a measure of error.
To obtain solutions, we require well-posedness, which necessitates information for six of the nine unknown parameters.
Using the~\citeauthor{Backus1962} average on a region of interest of a well log, we reduce the number of unknowns to four.
In particular, we specify $h_{1}$ and $h_{2}$\,, and calculate values for $\gamma=:\gamma^{\overline{\rm TI}}\,,\delta=:\delta^{\overline{\rm TI}}\,,\,\varepsilon=:\varepsilon^{\overline{\rm TI}}$\,, where~${}^{\overline{\rm TI}}$ denotes a quantity obtained using the~\citet{Backus1962} average.
Information for any of~$a_{S}\,,b_{S}\,,a_{P}\,,b_{P}$ must be obtained using additional methods.

To obtain information for the remaining model parameters, we use 1-D tomography and the~$ab$ model.
\subsection{1-D tomography}
\label{sec:1D_tomo}
To obtain linear inhomogeneity parameters from VSP data, we use a 1-D tomography method.\,
The inversion algorithm, therein, is based on the Levenberg-Marquardt least-square solution.
As a result of inversion, we obtain a velocity profile corresponding to discrete depths. 
The inverted velocity profile represents the velocity of a vertically inhomogeneous medium. 
To calculate the linear inhomogeneity parameters for a particular region, we apply the linear regression to the resultant velocities for the required depths.
The 1-D tomography can provide relatively good results in the case of a better initial model, a sufficient number of data, and that the assumption of linear inhomogeneity holds firmly for the medium.
\subsection{$\boldsymbol{ab}$ model}
The~$ab$ model is a traveltime inversion method.\,
In this method, the velocity model is a linear function of depth.  
To obtain the traveltime in a depth segment from borehole velocity, we assume the medium is composed of multiple thin isotropic layers.
Based on that assumption, we calculate Fermat's traveltime. 
In general, the borehole data contains high-frequency contents, which restricts the ray to travel from a source to a receiver with high offset.
To synthesize sufficient traveltime data, we need traveltime for both near and high offsets. 
That is possible only if we remove the higher frequencies from the data.
We apply a weighted average in every twelve consecutive data points to filter out those contents. 
As an outcome, we get a smoother velocity, which allows rays to travel from source to receiver with sufficiently more significant offset. 
In the inversion, we minimize the difference between the Fermat's traveltime and the model traveltime~\citep{SlawinskiSlawinski1999} and obtain linear inhomogeneity parameters.
\subsection{Mizzen O-16}
The VSP and well log data used, herein, was obtained from the Mizzen O-16 discovery well, which is a site in the Flemish Pass basin and was drilled in 2009 by Statoil~\citep{Enachescu2011}.
The well log data used is supplied by IHS energy; the data description is provided in~\citet{NalcorEnergy2016}.
We use $P$- and $S$-wavespeed measurements for depths of at depth 1865m to 2648.60\,m\,. 

The checkshot (VSP) data is provided by the Canada-Newfoundland \& Labrador Offshore Petroleum Board~\citep{CNLPOB2009}.
Therein, the traveltime data corresponds to a single source and multiple receivers. 
The source is located at a 26.5\,m offset and the receivers are placed along vertical axis, starting at depth 1865\,m and ending at 2650\,m. 
We consider our last receiver at depth 2650m so that the velocity inversion from traveltime merges with the region where the well log data are recorded.
The descriptions of the data are given in Tables~\ref{tab:vel_des} and \ref{tab:cs_des}, which are collected with the permission of the Petroleum Development Section of Natural Resources, Government of Newfoundland and Labrador. 

\begin{table}[h]
\centering
\begin{tabular}{c*{6}{c}}
\toprule
Field &
Well &
KB  (${\rm m}$) &
TVD  (${\rm m}$) &
Water depth (${\rm m}$) &
Spud date &
Log data
\\[2pt]
\toprule
Mizzen & O-16 & 21.15 & 3797 &1095 & 2008 & $\checkmark$ \\
\bottomrule
\end{tabular} 
\caption{
Description of the well for log data
}
\label{tab:vel_des}
\end{table}

\begin{table}[h]
\centering
\begin{tabular}{c*{6}{c}}
\toprule
Field &
Well &
Offset  (${\rm m}$) &
TVD  (${\rm m}$) &
Water depth (${\rm m}$) &
Spud date &
Checkshot
\\[2pt]
\toprule
Mizzen & O-16 & 26.50 & 3797 &1095 & 2009 & $\checkmark$ \\
\bottomrule
\end{tabular} 
\caption{
Description of the well for checkshot data
}
\label{tab:cs_des}
\end{table}
\section{Numerical search}
\label{sec:Results}
\subsection{Restrictions}
In our numerical search, we have two restrictions. 
First, the stability conditions of isotropy restrict the values for~$a_{S}$ and~$a_{P}$ such that
\begin{equation*}
a_{P} > 2\,a_{S}\,/\sqrt{3}
\,.
\end{equation*}
Second, to remain consistent with the assumption of constantly increasing velocity gradient with depth of~\citet{SlawinskiSlawinski1999}, we restrict our solutions to positive values of~$b_{P}$ and~$b_{S}$\,.
However, it is worth noting that there do exist solutions for negative values of~$b_{P}$ and~$b_{S}$\,.

\begin{figure}
\centering
\includegraphics[width=0.45\textwidth]{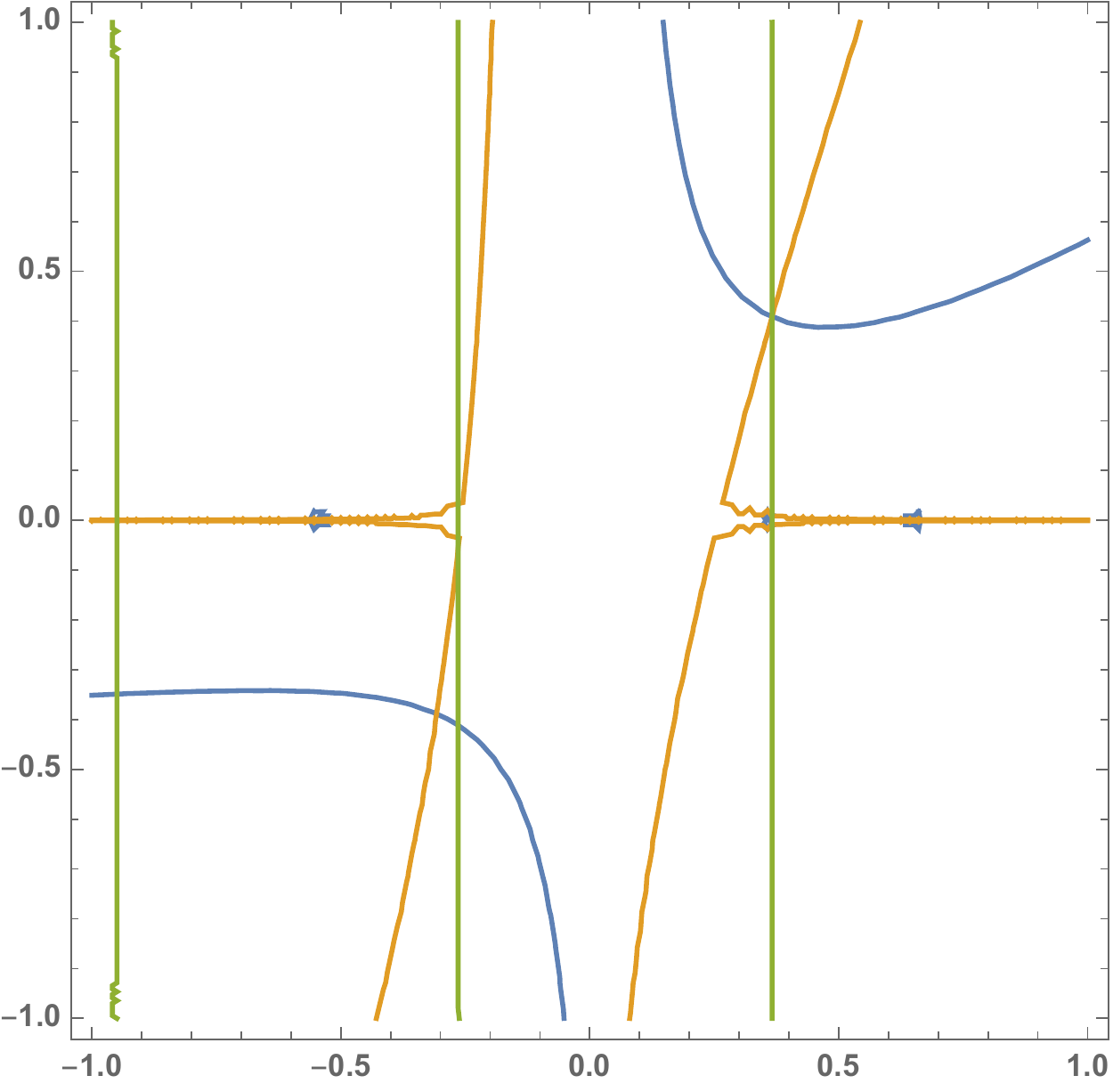}
\hspace*{\fill}
\includegraphics[width=0.45\textwidth]{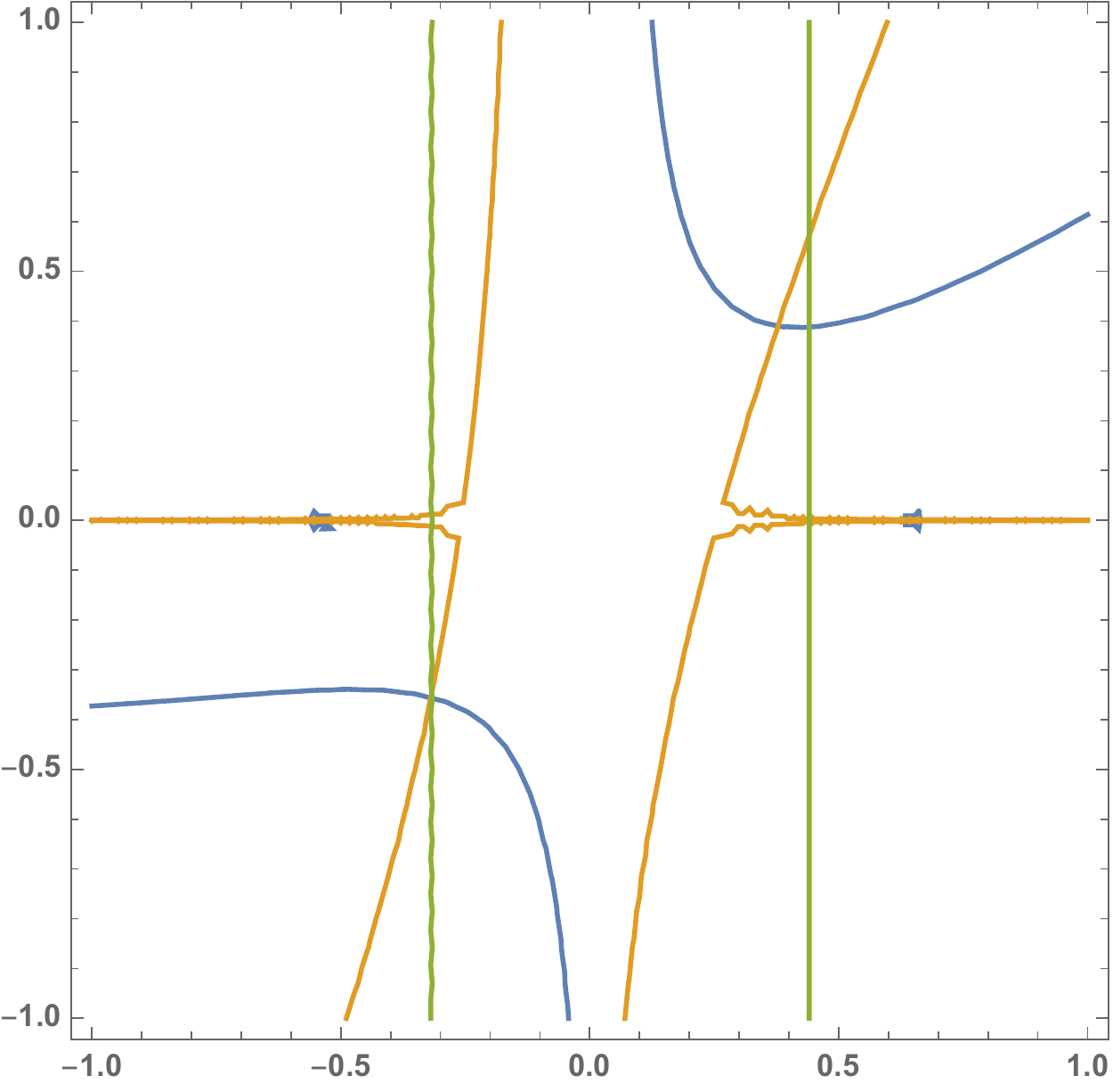}
\caption{%
	Contour plot of anisotropy values~\eqref{eq:AnisotropyValues}, where $b_{P}$ is along the vertical axis and $b_{S}$ is along the horizontal axis.
	Green lines represent $\gamma = \gamma^{\overline{\rm TI}}$\,; orange lines represents $\delta = \delta^{\overline{\rm TI}}$\,, blue lines represent $\varepsilon = \varepsilon^{\overline{\rm TI}}$\,.}
\label{fig:TriplePoints}
\end{figure}

We may demonstrate this with a numerical example.
For, say, an input of~$a_{P} = 2040.36\,{\rm ms}^{-1}$\,, there exist two solutions, as illustrated by the two instances of triple-point intersection in the left- and right-hand plots of Figure~\ref{fig:TriplePoints}.
Therein, the left-hand plot corresponds to a solution of positive~$b_{P}$ and~$b_{S}$\,, where
\begin{equation*}
\label{eq:VSP_anal_given_bp}
a_{S} = 752.95\,{\rm ms}^{-1}
\,,\quad
b_{S} = 0.3666\,{\rm s}^{-1}
\,,\quad
a_{P} = 2164.68\,{\rm ms}^{-1}
\,,\quad
b_{P} = 0.4081\,{\rm s}^{-1}
\,.
\end{equation*}
However, the right-hand plot corresponds to a solution of negative~$b_{P}$ and~$b_{S}$\,, where
\begin{equation*}
\label{eq:VSP_anal_given_bp}
a_{S} = 906.32\,{\rm ms}^{-1}
\,,\quad
b_{S} = -0.3194\,{\rm s}^{-1}
\,,\quad
a_{P} = 2164.68\,{\rm ms}^{-1}
\,,\quad
b_{P} = -0.3556\,{\rm s}^{-1}
\,.
\end{equation*}
Throughout the entirety of the numerical results, we consider positive solutions only.

\subsection{Calculations}
\label{sec:Calcs}
Let us obtain a solution to system~\eqref{eq:AnalyticSystem} using the methodology of Section~\ref{sec:1D_tomo} with the VSP data of Mizzen O-16.
Using results from 1-D tomography on traveltime data, we recover a velocity profile as a function of depth.
For a region of interest, whose depth ranges from 1865\,m to 2648.60\,m\,, we recover two parameters, namely~$a_{P}$ and~$b_{P}$\,.
We may use either parameter as input to system~\eqref{eq:AnalyticSystem} to obtain values for the remaining three unknowns.

For example, we use~$a_{P_{\rm VSP,in}} = 1415\,{\rm ms}^{-1}$ and~$b_{P_{\rm VSP,in}} = 0.70\,{\rm s}^{-1}$ as startup values for the tomography.
For the region of interest, we recover~$a_{P_{\rm VSP,out}} = 1354.9\,{\rm ms}^{-1}$ and~$b_{P_{\rm VSP,out}} = 0.3933\,{\rm s}^{-1}$\,, where, at the top of the region,~$a_{P_{\rm layer}} = a_{P_{\rm VSP,out}} + b_{P_{\rm VSP,out}}\cdot1865\,{\rm m} = 2088.38\,{\rm ms}^{-1}$\,.
We input~$b_{P} = b_{P_{\rm VSP,out}}$ into system~\eqref{eq:AnalyticSystem}, for 
\begin{equation*}
h_{1} = 0\,{\rm m}
\,,\quad
h_{2} = \left(2648.6 - 1865\right){\rm m} = 783.6\,{\rm m}\,,
\end{equation*}
\begin{equation}
\label{eq:AnisotropyValues}
\gamma^{\overline{\rm TI}} = 0.017561151400350
\,,\quad
\delta^{\overline{\rm TI}} = -0.005822848520484
\,,\quad
\varepsilon^{\overline{\rm TI}} = 0.002868244418444
\,,
\end{equation}
and obtain
\begin{equation}
\label{eq:VSP_anal_given_bp}
a_{S} = 725.55\,{\rm ms}^{-1}
\,,\quad
b_{S} = 0.3533\,{\rm s}^{-1}
\,,\quad
a_{P} = 2085.91\,{\rm ms}^{-1}
\,,\quad
b_{P} = 0.3933\,{\rm s}^{-1}
\,.
\end{equation}
It is clear that~$a_{P_{\rm layer}}\neq a_{P}$ but we may perform an error analysis to assess the ``closeness'' of our result.
To do so, we recall expressions~\eqref{eq:gammaDifferential}, \eqref{eq:deltaDifferential}, \eqref{eq:epsilonDifferential}, which are total differentials ${\rm d}\gamma\,,{\rm d}\delta\,,{\rm d}\varepsilon$\,.
Using uncertainty measures for parameters~\eqref{eq:VSP_anal_given_bp}, where
\begin{equation}
\label{eq:Aniso_Unc_param}
{\rm d}h_{1} = {\rm d}h_{2} = 0.05\,{\rm m}
\,,\quad
{\rm d}a_{S} = {\rm d}a_{P} = 2\,{\rm ms}^{-1}
\,,\quad
{\rm d}b_{S} = {\rm d}b_{P} = 0.01\,{\rm s}^{-1}
\,,
\end{equation}
we obtain
\begin{equation}
\label{eq:anisotropy_differentials}
{\rm d}\gamma = 0.000790010132156
\,,\quad
{\rm d}\delta = -0.000299948944243
\,,\quad
{\rm d}\varepsilon = 0.000143889902574
\,.
\end{equation}
We obtain a new set of solutions for the lower limit of anisotropy by subtracting values~\eqref{eq:anisotropy_differentials} from the left-hand sides of system~\eqref{eq:AnalyticSystem}, which are
\begin{equation}
\label{eq:VSP_anal_given_bp_sub}
a_{S_{-}} = 742.47\,{\rm ms}^{-1}
\,,\quad
b_{S_{-}} = 0.3522\,{\rm s}^{-1}
\,,\quad
a_{P_{-}} = 2138.75\,{\rm ms}^{-1}
\,,\quad
b_{P} = 0.3933\,{\rm s}^{-1}
\,.
\end{equation}
Similarly, the upper limit of anisotropy is obtained adding values~\eqref{eq:anisotropy_differentials}, which results in
\begin{equation}
\label{eq:VSP_anal_given_bp_add}
a_{S_{+}} = 709.58\,{\rm ms}^{-1}
\,,\quad
b_{S_{+}} = 0.354246\,{\rm s}^{-1}
\,,\quad
a_{P_{+}} = 2036.58\,{\rm ms}^{-1}
\,,\quad
b_{P} = 0.3933\,{\rm s}^{-1}
\,.
\end{equation}
Thus, we find that~$a_{P_{\rm layer}}$ falls within the range determined by~$a_{P_{-}}$ and~$a_{P_{+}}$\,.
In particular, such startup values provide results that are within 0.12\% and 0.13\%\,, respectively, of the theoretical predictions.
We may repeat this entire process using~$a_{P} = a_{P_{\rm VSP,out}}$ as input.
In addition, this process may be repeated for a range of VSP startup values, in order to find the combination of startup values that are closest to the predicted model parameters indicated by the analytical relation.
We tabulate the results of such a process, for values of~$a_{P_{\rm VSP,in}}$ ranging from 1325\,ms$^{-1}$ to 1515\,ms$^{-1}$\,, with increments of 20\,ms$^{-1}$\,, and~$b_{P_{\rm VSP,in}}$ ranging from 0.57\,s$^{-1}$ to 0.7\,s$^{-1}$\,, with increments of 0.01\,s$^{-1}$\,, in Table~\ref{tab:VSP_results} of Appendix~\ref{app:Tables}.

To compare the results from VSP with the~$ab$ model, we repeat this process for the same range of values of~$a_{P_{\rm ab,in}}$ and~$b_{P_{\rm ab,in}}$\,.
The results are tabulated in Table~\ref{tab:ab_results} of Appendix~\ref{app:Tables}.
In comparison, the startup values of 
~$a_{P_{ab,{\rm in}}} = 1625\,{\rm ms}^{-1}$ and~$b_{P_{ab,{\rm in}}} = 0.6\,{\rm s}^{-1}$, provide results that are within 0.03\% and 0.03\%\,, respectively, of the theoretical predictions.

We illustrate the entirety of the results of Tables~\ref{tab:VSP_results} and~\ref{tab:ab_results} in Figures~\ref{fig:fig_bp_given_ap_new} and~\ref{fig:fig_ap_given_bp_new}.
Therein, the solid red line represents the solution to system~\eqref{eq:AnalyticSystem} corresponding to the value on the horizontal axis, and the solid black lines represent solutions for the lower and upper limits of anisotropy, corresponding to uncertainty parameters~\eqref{eq:Aniso_Unc_param}.
Orange diamonds represent the solutions from 1-D tomography and blue dots represent solutions from the~$ab$ model.
\section{Discussion}
\begin{figure}[h]
\centering
\subcaptionbox{\label{fig:fig_bp_given_ap_new}}{\includegraphics[width=0.49\textwidth]{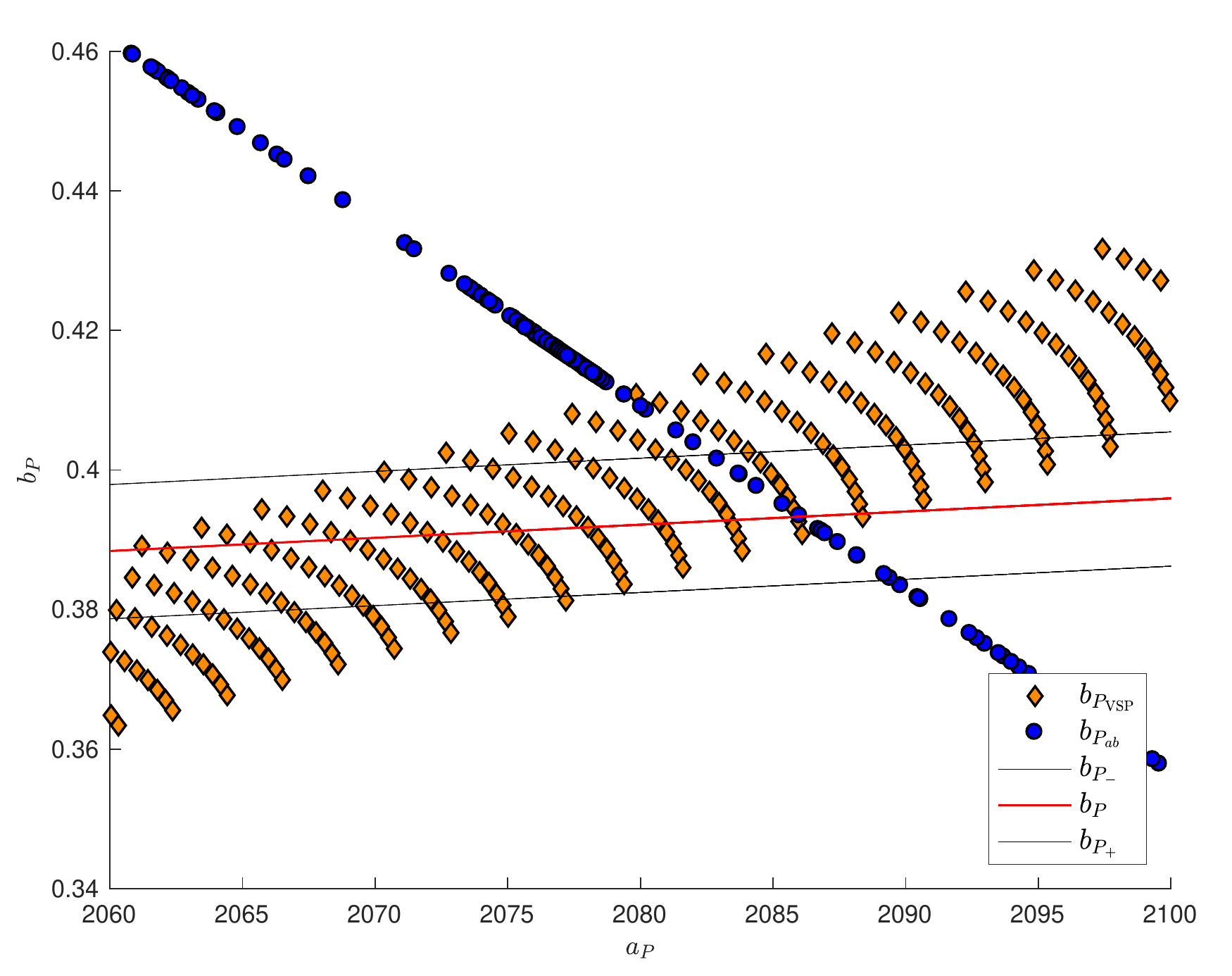}}\hfill
\subcaptionbox{\label{fig:fig_ap_given_bp_new}}{\includegraphics[width=0.49\textwidth]{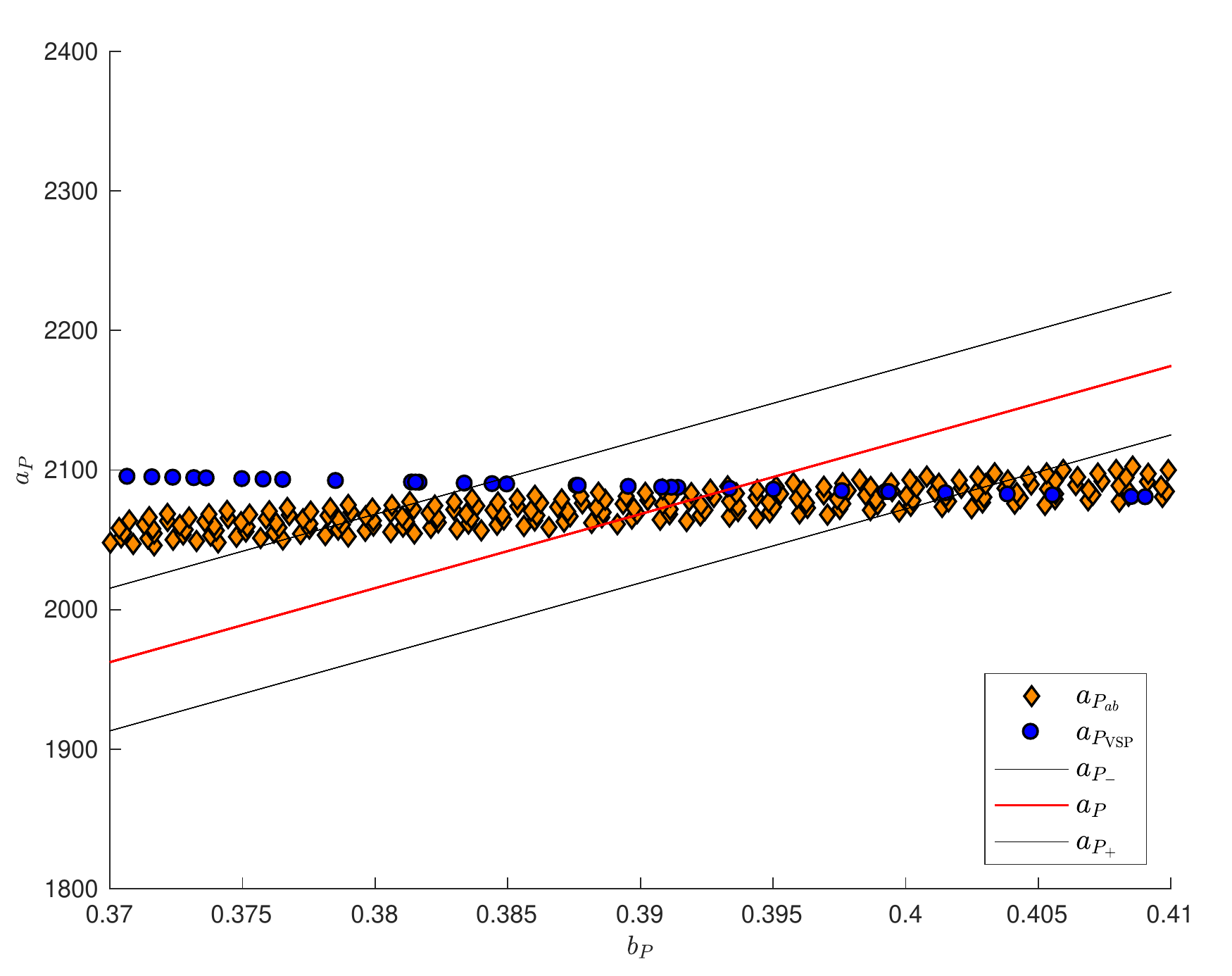}}
\caption{%
Solutions of 1-D tomography and~$ab$-model for startup values of $a_{P_{\rm VSP,in}}$ and~$b_{P_{\rm VSP,in}}$ in Table with VSP results and $a_{P_{ab,{\rm in}}}$ and~$b_{P_{ab,{\rm in}}}$ in Table with $ab$ results.
}
\label{fig:fig_bp_given_ap_ap_given_bp_new}
\end{figure}

For each solution method, we obtain some $a_{P_{{\rm VSP},ab}}$ and $b_{P_{{\rm VSP},ab}}$ parameter.
Using the outputted~$a_{P_{{\rm VSP},ab}}$ as input to system~\eqref{eq:AnalyticSystem}, we calculate~$b_{P}$\,,~$b_{P_{-}}$\,,~$b_{P_{+}}$\,, and compare them to the outputted $b_{P_{{\rm VSP},ab}}$\,. 
These results are illustrated in Figure~\ref{fig:fig_bp_given_ap_new}; the opposite operation is performed and those results are illustrated in Figure~\ref{fig:fig_ap_given_bp_new}.

\begin{figure}[h]
\centering
\subcaptionbox{\label{fig:fig_bp_given_ap_solution}}{\includegraphics[width=0.49\textwidth]{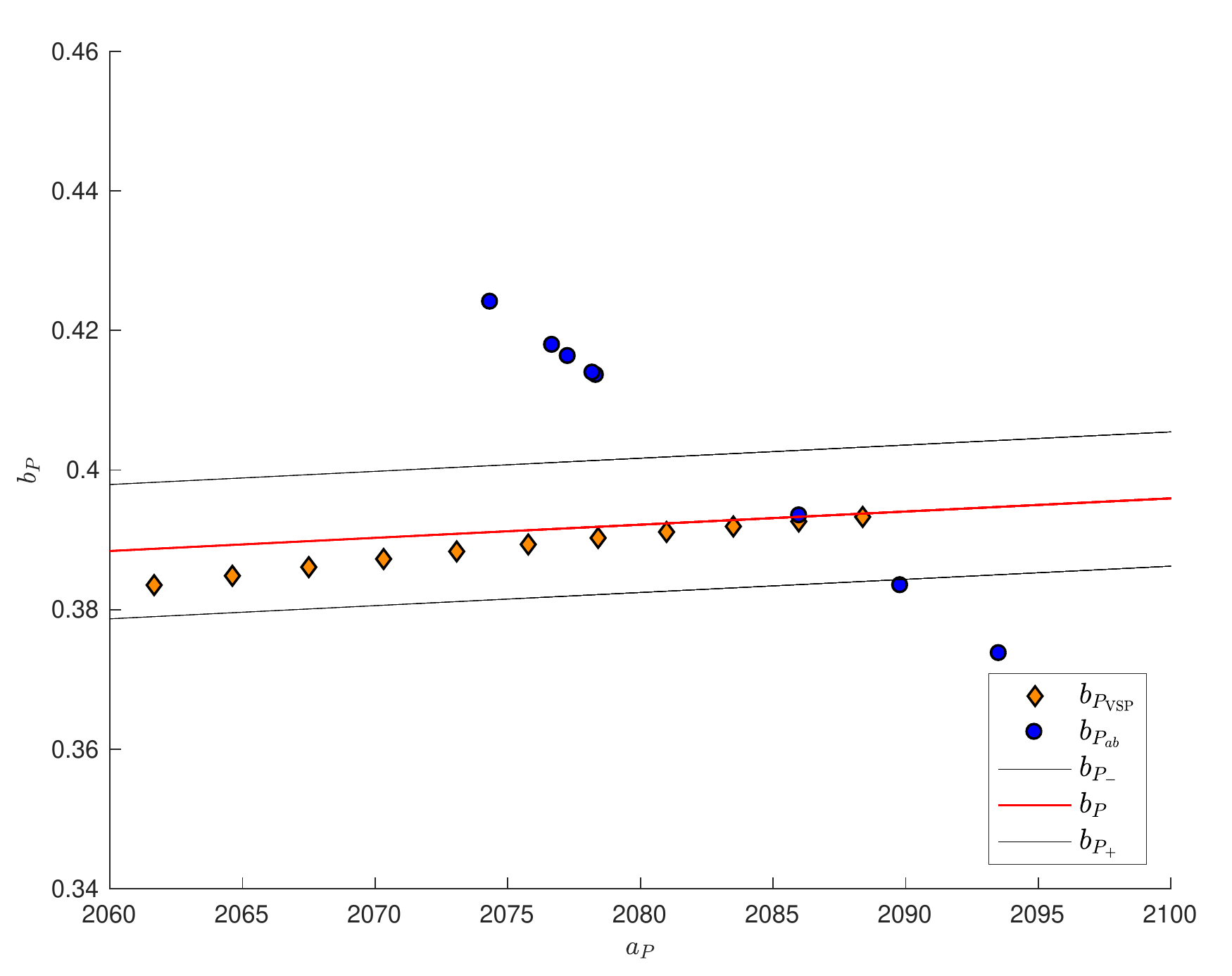}}\hfill
\subcaptionbox{\label{fig:fig_ap_given_bp_solution}}{\includegraphics[width=0.49\textwidth]{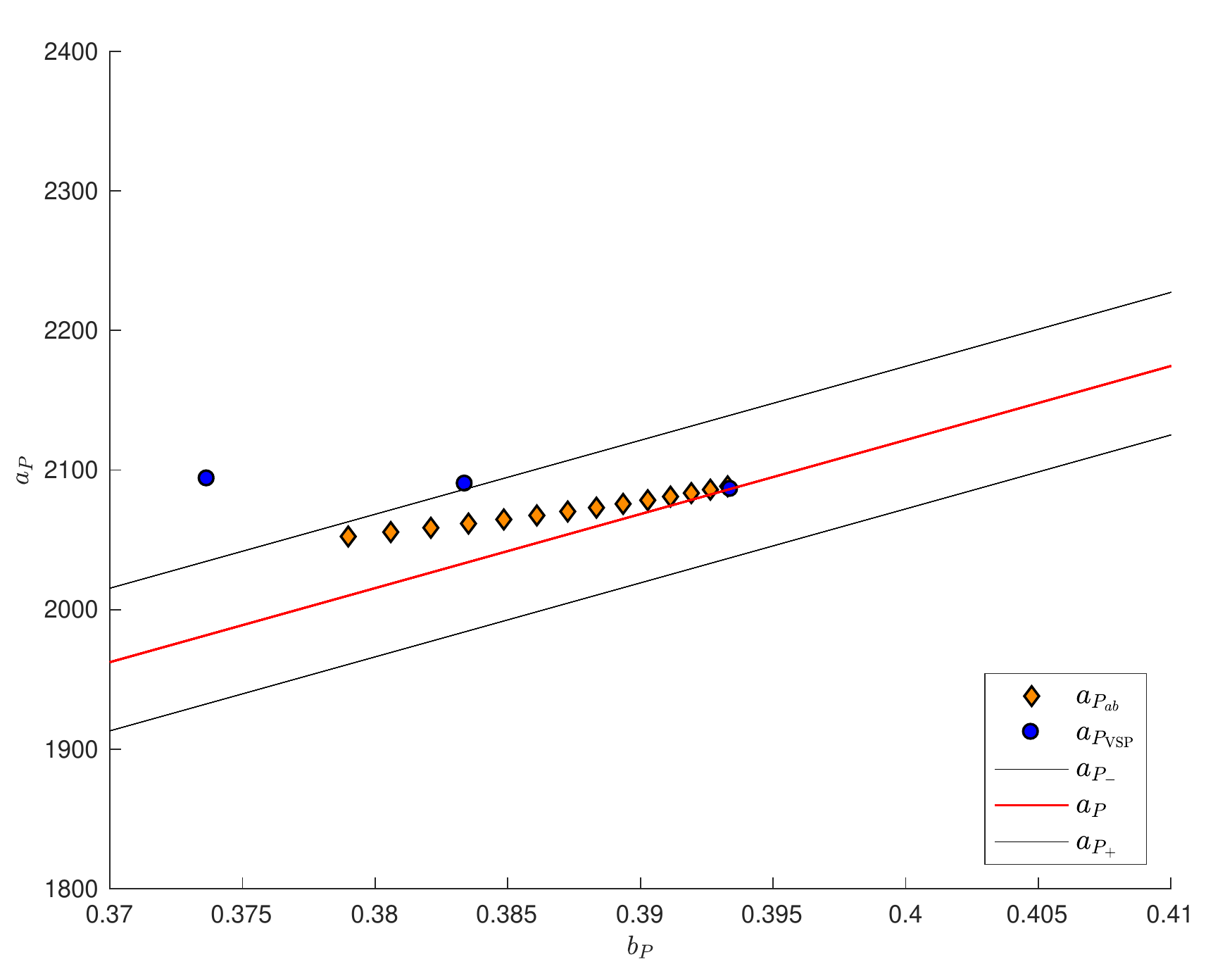}}
\caption{%
Solutions of 1-D tomography and~$ab$-model for startup values of $a_{P_{\rm VSP,in}}=1415$\,ms$^{-1}$ and~$b_{P_{\rm VSP,in}}$ incrementing by 0.01\,s$^{-1}$ from 0.57\,s$^{-1}$ to 0.7\,s$^{-1}$\,, and $a_{P_{ab,{\rm in}}} = 1625$\,ms$^{-1}$ and~$b_{P_{ab,{\rm in}}}$ incrementing by 0.05\,s$^{-1}$ from 0.25\,s$^{-1}$ to 0.65\,s$^{-1}$.
}
\label{fig:fig_bp_given_ap_ap_given_bp_solution}
\end{figure}

In both subplots of Figure~\ref{fig:fig_bp_given_ap_ap_given_bp_new}, we observe that the area of solutions for 1-D tomography overlaps the line of~$ab$-model solutions.
To discern which startup values lead to common outputs, we turn our attention to Figure~\ref{fig:fig_bp_given_ap_ap_given_bp_solution}.
Therein, startup values of $a_{P_{{\rm VSP,in}}} = 1415$\,ms$^{-1}$ and $b_{P_{{\rm VSP,in}}} = 0.7$\,ms$^{-1}$ result in$a_{P_{{\rm VSP,out}}} = 2088.38$\,ms$^{-1}$ and $b_{P_{{\rm VSP,out}}} = 0.3932$\,ms$^{-1}$\,.
These results, as stated in Section~\ref{sec:Calcs}, are within 0.12\% and 0.13\%\,, respectively, of the theoretical predictions.
Similarly, startup values of $a_{P_{ab,in}} = 1625$\,ms$^{-1}$ and $b_{P_{ab,in}} = 0.6$\,ms$^{-1}$ result in outputs that are within 0.03\% of the theoretical predictions.

\section{Conclusion and future work}
\label{sec:Conclusion}
In this article, we show that we can obtain multiple solution from VSP and $ab$-model that do not lie in the theoretical range.
However, we also show that there exist some common solution. 

Based on the initial model, we find that there are some initial values that lead to that common solution, which may allow us to conclude that, by tuning initial model parameters, we can obtain the common solution set.

The number of traveltime data from the field measurement plays an important role in the VSP method.
As a future work, we plan to apply the analytical relation and seismic methods to a site that possesses more data, which may produces a stable solution for different initial models in the VSP method. 

By far, we have shown that the relationship between inhomogeneity and anisotropy provides reasonable solution in compare to the seismic methods. 
In addition to our future project, we wish to examine the limitation of the analytical relationship by calculating Backus average from a sparse well log data.
\bibliographystyle{apa}
\bibliography{AS.bib}
\appendix
\section{Tables}
\label{app:Tables}
{\tiny
\begin{longtable}{c*{11}{c}}
\caption{VSP results}\label{tab:VSP_results}\\
\multicolumn{2}{c}{Startup values} & 
\multicolumn{2}{c}{VSP results} &
& \multicolumn{3}{c}{Using $b_{P_{\rm VSP,out}}$ as input} &
& \multicolumn{3}{c}{Using $a_{P_{\rm layer}}$ as input} \\
\toprule
$a_{P_{\rm VSP,in}}$ & $b_{P_{\rm VSP,in}}$ & $a_{P_{\rm VSP,out}}$ & $b_{P_{\rm VSP,out}}$ &
$a_{P_{\rm layer}}$ & $a_{P_{+}}$ & $a_{P}$ & $a_{P_{-}}$ & 
$b_{P_{\rm VSP,out}}$ & $b_{P_{-}}$ & $b_{P}$ & $b_{P_{+}}$ \\[2pt]
\toprule
1320.0 & 0.57 & 1292.3 & 0.4317 & 2097.42 & 2240.18 & 2289.65 & 2342.32 & 0.4317 & 0.3857 & 0.3955 & 0.405\\ 1325.0 & 0.57 & 1295.46 & 0.4286 & 2094.83 & 2223.84 & 2273.3 & 2325.98 & 0.4286 & 0.3853 & 0.395 & 0.4045\\ 1330.0 & 0.57 & 1298.58 & 0.4256 & 2092.27 & 2207.69 & 2257.14 & 2309.83 & 0.4256 & 0.3848 & 0.3945 & 0.404\\ 1335.0 & 0.57 & 1301.65 & 0.4226 & 2089.73 & 2191.74 & 2241.18 & 2293.89 & 0.4226 & 0.3843 & 0.394 & 0.4035\\ 1340.0 & 0.57 & 1304.68 & 0.4196 & 2087.22 & 2175.98 & 2225.41 & 2278.13 & 0.4196 & 0.3838 & 0.3935 & 0.4031\\
\bottomrule
\end{longtable}
}

{\tiny
\begin{longtable}{c*{9}{c}}
\caption{$ab$-model results}\label{tab:ab_results}\\
\multicolumn{2}{c}{Startup values} & &
\multicolumn{3}{c}{Using $b_{P_{ab,{\rm out}}}$ as input} &
& \multicolumn{3}{c}{Using $a_{P_{ab,{\rm out}}}$ as input} \\
\toprule
$a_{P_{ab,{\rm in}}}$ & $b_{P_{ab,{\rm in}}}$ &
$a_{P_{ab,{\rm out}}}$ & $a_{P_{+}}$ & $a_{P}$ & $a_{P_{-}}$ & 
$b_{P_{ab,{\rm out}}}$ & $b_{P_{-}}$ & $b_{P}$ & $b_{P_{+}}$ \\[2pt]
\toprule
1005.0 & 0.25 & 2078.35 & 2246.06 & 2193.31 & 2143.9 & 0.4135 & 0.4014 & 0.3919 & 0.3821\\ 1025.0 & 0.25 & 2078.35 & 2246.05 & 2193.3 & 2143.89 & 0.4135 & 0.4014 & 0.3919 & 0.3821\\ 1045.0 & 0.25 & 2077.9 & 2252.09 & 2199.35 & 2149.94 & 0.4147 & 0.4013 & 0.3918 & 0.3821\\ 1065.0 & 0.25 & 2078.04 & 2250.26 & 2197.51 & 2148.1 & 0.4143 & 0.4013 & 0.3918 & 0.3821\\ 1085.0 & 0.25 & 2078.14 & 2248.84 & 2196.1 & 2146.69 & 0.4141 & 0.4013 & 0.3918 & 0.3821\\
\bottomrule
\end{longtable}
}
\end{document}